\documentclass{icsm}
\usepackage{epsfig}

\slacs{.4ex}

\begin{document}
\title{From the Mendeleev periodic table to 
particle physics \\ 
and back to the periodic table}
\authori{Maurice R. Kibler}
\addressi{Universit\'e de Lyon, Institut de Physique Nucl\'eaire,
          universit\'e Lyon~1 and CNRS/IN2P3 \\
43 Bd du 11 Novembre 1918, F--69622 Villeurbanne Cedex, France}
\authorii{}     \addressii{}
\authoriii{}    \addressiii{}
\authoriv{}     \addressiv{}
\authorv{}      \addressv{}
\authorvi{}     \addressvi{}
\headauthor{M.R. Kibler}
\headtitle{From the Mendeleev periodic table \ldots}

\lastevenhead{M.R. Kibler: From the Mendeleev periodic table \ldots}

\maketitle

\begin{abstract}
We briefly describe in this paper the passage from Mendeleev's 
chemistry (1869) to atomic physics (in the 1900's), nuclear 
physics (in 1932) and particle physics (from 1953 to 2006). 
We show how the consideration of symmetries, largely used in physics 
since the end of the 1920's, gave rise to a new format of the periodic 
table in the 1970's. More specifically, this paper is concerned with 
the application of the group SO(4,2)$\otimes$SU(2) to the periodic 
table of chemical elements. It is shown how the Madelung rule of the 
atomic shell model can be used to set up a periodic table that 
can be further rationalized via the group SO(4,2)$\otimes$SU(2) and 
some of its subgroups. Qualitative results are obtained from this 
nonstandard table. 
\end{abstract}

\section{Introduction}

Antoine Laurent de Lavoisier (1743-1794) is certainly the father 
of modern chemistry. Nevertheless, the first important progress 
in the classification of chemical elements is due to Dimitri 
Ivanovitch Mendeleev (1834-1907). Indeed, the classification 
of elements did not start with Mendeleev. The impact for 
chemistry of numerous precursors of Men\-de\-leev is well--known 
(van Spronsen, 1969; Rouvray and King, 2004; Scerri, 2007). For 
example we can mention, 
among others, Johann W. D\"obereiner and his triads of elements (1829), 
Max von Pettenkofer and his groupings of elements (1850), Alexandre E. 
Beguyer de Chancourtois and his spiral or telluric periodic table 
(1862), John A.R Newlands and his law of octaves (1864), William Odling and 
his periodic table (1864), and Julius Lothar Meyer and his curve of 
atomic volumes (1868). In spite of the interest of the works of his 
predecessors, Mendeleev is recognized as the originator of the 
classification of elements for the following reasons. As a matter 
of fact, in 1869 Mendeleev was successful in four directions: 
(i) he gave a classification of elements according to growing 
atomic weights; (ii) he made two inversions (Te/I and Ni/Co) violating 
the ordering via growing atomic weights and thus compatible 
with the now accepted ordering via growing atomic numbers; 
(iii) he predicted 
the existence of new elements (via the eka-process); and (iv) 
he described in a qualitative and quantitative way the main 
chemical and physical properties of the predicted elements. It 
is true that other scientists tried to develop ideas along the 
lines (i)-(iv). However, Mendeleev was the first to make 
observable predictions for new elements. A question naturally 
arises: What happened after the establishment of the 
Mendeleev periodic table?" There are two answers.

First, let us mention the development and the extension of the table 
with the discovery of new elements. For instance, the element called 
eka-silicon predicted by Mendeleev was discovered 
in 1886. Such an element, now called germanium, belongs to the same 
column as silicon (it is located below silicon and above tin 
in the periodic table with horizontal lines). This discovery illustrates 
the `eka'-process or `something is missing'-process: In order to respect 
some regularity and periodicity arguments, Mendeleev left an empty 
box at the right of silicon (in his periodic table with vertical 
lines) and predicted a new element with the correct mass and density.

The discovery of new elements continued during the end of the 19th 
century, during the whole 20th century and is still the subject of 
experimental and theoretical investigations. Approximately, 70 elements 
were known in 1870, 86 in 1940 and 102 in 1958. In the present days, 
we have 116 elements 
(the last ones are less and less stable). Some of the recently observed 
elements do not have a name (the last named is called roentgenium). There 
is no major reason, except for experimental reasons, to have an end for 
the periodic table. The research for heavy elements (indeed, superheavy nuclei) 
is far from being finished.

Second, let us mention a spectacular advance in the understanding of the 
complexity of matter. This gave rise to the discovery of sub-structures 
with the advent of classification tables for the constituents of the chemical 
elements themselves. More precisely, the discovery of atomic structure 
led to atomic physics at the beginning of the 20th century, then to the birth 
of nuclear physics in 1932 and, finally, to particle physics in the 1950's. In 
the present days, subatomic physics deals with the elementary constituents 
of matter and with the forces or interactions between these constituents. 

\section{From chemistry to atomic, nuclear and particle physics}

According to Mendeleev, a chemical element had no internal structure. The 
chemical elements are made of atoms without constituents. The 
discovery of the electron by J.J. Thomson in 1897 and that of the atomic 
nucleus by E. Rutherford in 1911 led to the idea of a planetary model for 
the atom where the electrons orbit around the nucleus. The simplest nucleus, 
namely, the proton, was observed by Rutherford in 1919. The simplest atom, the 
hydrogen atom, is made of a nucleus consisting of a single proton considered as 
fixed and of an electron moving around the proton. Atomic spectroscopy, seen 
via the prism of the old theory of quanta (the starting point for quantum 
mechanics), was born in 1913 when N. Bohr introduced his semi-classical 
treatment of the hydrogen atom. In 1922, Bohr proposed a building-up principle 
for the atom based on the planetary Bohr-Sommerfeld model with elliptic orbits 
and on the filling of each orbit with a maximum of two electrons. This led him 
to adopt a pyramidal form for the periodic table, that had been already 
proposed by others like Bayley, and to predict that hafnium is a transition 
metal as opposed to a rare earth (Scerri, 1994)."    

A further step towards complexity occurred with the discovery of the neutron by 
J. Chadwick in 1932. A nucleus is made of protons and neutrons, collectively 
denoted as nucleons. A proton has a positive electric charge, which is the 
opposite of the electronic charge of the electron, 
and the neutron has no electric charge. The 
discovery of a substructure for the nucleus opened the door of nuclear physics. 
The first model for the description of the strong interactions between nucleons 
inside the nucleus, the SU(2) model of W. Heisenberg, goes back to 1932. It was soon 
completed by the prediction by I. Yukawa of the meson $\pi$~in 1933. It 
can be said that the first version of the strong interaction theory (the 
ancestor of quantum chromodynamics developed in the 1970's) was born with the 
works of Heisenberg and Yukawa. In a parallel way, E. Fermi developed in 
1933 a theory for the weak interactions inside the nucleus (the ancestor of 
the electroweak model developed in the 1960's).

In 1932, the situation for emerging particle physics was very simple and 
symmetrical. At this time, we had four particles: two hadrons (proton and 
neutron) and two leptons (electron and neutrino), as well as their 
antiparticles resulting from the P.A.M. Dirac relativistic quantum mechanics 
introduced in 1928. Remember that the neutrino (in fact the antineutrino of 
the electron) was postulated by W. Pauli in 1931 in order to ensure the 
conservation of energy. In some sense, the introduction of the neutrino was 
made along lines that parallel the eka-process used by Mendeleev. Furthermore, 
with the discovery of a new lepton, the muon in 1937, of three new hadrons, 
the pions in 1947-50, and of a cascade of strange hadronic particles (mesons 
and baryons) in cosmic rays in the 1950's, particle physics was in the 1950's 
in a situation similar to the one experienced by chemistry in the 1860's. The 
need for a classification was in order. In this direction, S. Sakata tried 
without success to introduce in 1956 three elementary particles (proton, 
neutron, lambda particle) from which it would be possible to generate all 
hadrons. Indeed, this trial based on the group SU(3) was an extension of the 
model developed by E. Fermi and C.N. Yang in 1949, based on the group SU(2), 
with two basic hadrons (proton, neutron). 

A decisive step was made with the 
introduction of the so-called eightfold way by M. Gell-Mann and 
Y. Ne'eman in 1961. The eightfold way is a model for the classification of 
hadrons within multiplets (singlets, octets and decuplets) corresponding to 
some irreducible representation classes (IRCs) of the group SU(3). More 
precisely, the eight $(0)^-$ pseudo-scalar mesons and the eight $(1)^-$ vector 
mesons were classified in two nonets (nonet = `octet plus singlet') while the 
eight $(\frac{1}{2})^+$ baryons were classified 
in an octet. At this time, we knew nine 
$(\frac{3}{2})^+$ baryons (the four $\Delta$, the three $\Sigma^*$ and the two 
$\Xi^*$). It was not possible to accommodate these nine particles into an 
`octet plus singlet'. The closer framework or `perodic table'" for 
accomodating the nine hadrons was a decuplet with ten boxes. Along the lines 
of an eka-process, in 1962 Gell-Mann was very 
well inspired to fill the empty box with 
a new particle, the particle $\Omega$. He was also able to predict the main 
characteristics of this postulated particle 
(spin, parity, isospin, charge, mass, etc.). This 
particle was observed two years later, in 1964. The interest of 
group theory for classifying purposes was thus clearly established. The 
relevance and usefulness of SU(3) 
was confirmed with the introduction of new particles in 1964: 
the `quarks'"of M. Gell-Mann and the `aces'"of G. Zweig. Gell-Mann and Zweig 
postulated the existence of three elementary particles and their 
anti-particles, now called quarks and antiquarks, 
classified into a triplet and an anti-triplet of SU(3) from 
which it is possible to generate all hadrons. 

As soon as 1970, a fourth quark was 
postulated by S.L. Glashow, J. Iliopoulos 
and L. Maiani to explain the non-observation of certain {\it a priori} allowed 
decays. This quark was indirectly observed in 1974 through the production of a
charmed meson so that the matter world was 
again very symmetrical at that time with
four quarks ($u, d, c, s$) and four leptons ($e, \mu, \nu_e, \nu_{\mu}$). The
interest, for the classification of hadrons, of symmetry groups like SU($n$),
now called flavor groups, was then fully confirmed. Besides flavor groups, 
other groups, called gauge groups, appeared during the 1960's and 1970's to
describe interactions between particles. Let us mention the group 
SU(2)$\otimes$U(1) for the weak and electromagnetic interactions, the group
SU(3) for the strong interactions and the groups SU(5), SO(10) and E$_6$ for a 
grand unified description of electroweak and strong interactions. In addition,
the supersymmetric Poincar\'e group was introduced for unifying
external (space-time) symmetries and internal (flavor and gauge) symmetries. As
a result of investigations based on symmetries and supersymmetries, we now have 
in 2006 the standard model and its supersymmetric extensions for describing 
particles and their interactions (gravitation excluded). This model is based 
on twelve matter fields, twelve gauge fields 
mediating interactions between matter fields and one feeding particle 
(the Higgs boson) which gives mass to massive particles. Indeed, the matter
particles (six quarks and six leptons) can be accomotaded in a periodic table
with three generations or periods. 

The advances in group--theoretical methods in direction of particle physics
as well as the introduction of invariance groups and noninvariance groups
for describing dynamical systems were a source of inspiration for the 
use of groups in connection with the periodic table. The rest of this paper
is devoted to the building of a periodic table based on the direct product 
group SO(4,2)$\otimes$SU(2).

\section{Introducing the group SO(4,2)}

Most of the modern presentations of the periodic table of chemical
elements are based on a quantum--mechanical treatment of the atom.
In this respect, the simplest atom, namely the hydrogen atom,
often constitutes a starting point for studying many--electron
atoms. Naively, we may expect to construct an atom with atomic
number $Z$ by distributing $Z$ electrons on the one--electron
energy levels of a hydrogen--like atom. This building-up principle
can be rationalized and refined from a group--theoretical point of
view. As a matter of fact, we know that the dynamical
noninvariance group of a hydrogen--like atom is the special real
pseudo-orthogonal group in 4+2 dimensions 
SO(4,2) or SO(4,2)$\otimes$SU(2) if we introduce the group
SU(2) that labels the spin (Malkin and Man'ko, 1965; 
Barut and Kleinert, 1967). This 
result can be derived in several ways. We briefly review two of them 
(one is well--known, the other is little known).

The first way corresponds to a symmetry
ascent process starting from the geometrical symmetry group
SO(3) of a hydrogen--like atom. Then, we go from SO(3) to the
dynamical invariance group SO(4) for the discrete spectrum or
SO(3,1) for the continuous spectrum. The relevant quantum
numbers for the discrete spectrum are $n$, $\ell$ and $m_{\ell}$
(with $n=1,2,3,\cdots$; for fixed $n$: $\ell=0,1,\cdots,n-1$; for
fixed $\ell$: $m_{\ell}=-\ell,-\ell+1,\cdots,\ell$). The
corresponding state vectors $\Psi_{n\ell m_{\ell}}$ can be
organized to span multiplets of SO(3) and SO(4). The set
$\{\Psi_{n\ell m_{\ell}}:n{\rm~and~}\ell{\rm~fixed~};\,m_{\ell}
{\rm~ranging}\}$ generates an 
IRC of SO(3), noted ($\ell$), while the set $\{\Psi_{n \ell m_{\ell}}:n
{\rm~fixed}\,;\ell{\rm~and~}m_{\ell}{\rm~ranging}\}$ generates an
IRC of SO(4). The direct sum
$$
h=\bigoplus_{n=1}^{\infty}\bigoplus_{\ell=0}^{n-1}(\ell)
$$
spanned by all the possible state vectors $\Psi_{n\ell m_{\ell}}$
corresponds to an IRC of the de Sitter group SO(4,1). The IRC
$h$ is also an IRC of SO(4,2). This IRC thus remains irreducible
when restricting SO(4,2) to SO(4,1) but splits into two IRC's
when restricting SO(4,2) to SO(3,2). The groups SO(4,2),
SO(4,1) and SO(3,2) are dynamical noninvariance groups in the
sense that not all their generators commute with the Hamiltonian
of the hydrogen--like atom.

The second way to derive SO(4,2) corresponds to a symmetry
descent process starting from the dynamical noninvariance group
Sp(8,{\bf R}), the real symplectic group in 8 dimensions, for a
four--dimensional isotro\-pic harmonic oscillator. We know that
there is a connection between the hydrogen-like atom in ${\bf
R}^3$ and a four--dimensional oscillator 
in ${\bf R}^4$ (Kibler and N\'egadi, 1984). 
Such a connection can be established via Lie--like methods (local
or infinitesimal approach) or algebraic methods based on the
so-called Kustaanheimo--Stiefel transformation (global or partial
differential equation approach). Both approaches give rise to a
constraint and the introduction of this constraint into the Lie
algebra of Sp(8,{\bf R}) produces a Lie algebra under
constraints that turns out to be isomorphic with the Lie algebra
of SO(4,2). From a mathematical point of view, the latter Lie
algebra is given by 
$$
\mathrm{cent}_{{\rm sp}(8,{\bf R})}{\rm so}(2)/{\rm so}(2) = 
{\rm su}(2,2) \sim {\rm so}(4,2)
$$
in terms of Lie algebras.

Once we accept that the hydrogen--like atom may serve as a guide
for studying the periodic table, the group SO(4,2) and some of
its subgroups play an important role in the construction of this
table. This was first realized by Rumer and Fet (Rumer and Fet, 1971) 
and, independently, by Barut (Barut, 1972). Later, 
Byakov, Kulakov, Rumer and Fet (Konopel'chenko and Rumer, 1979)
further developed this group--theoretical approach of the periodic chart
of chemical elements by introducing the direct product
SO(4,2)$\otimes$SU(2) and Kibler (Kibler, 2004, 2006)
fully described the SO(4,2)$\otimes$SU(2) table in connection with 
the so-called Madelung rule of atomic spectroscopy. 

\section{The periodic table {\bf\textit{\`a la}} Madelung}

Before introducing the table based on SO(4,2)$\otimes$SU(2), 
we describe the construction of a periodic table based on
the Madelung rule which arises from the atomic shell 
model. This approach to the periodic table uses the quantum
numbers occurring in the quantum--mechanical treatment of the
hydrogen atom as well as of a many--electron atom.

Several authors have claimed that the Madelung rule has not 
been deduced from the first principles of Quantum Mechanics 
(L\"owdin, 1969; Scerri, 2006). This is certainly true if we 
limit the study of quantum dynamical systems to the paradigmatic 
quantum systems (and their trivial extensions), namely, the Coulomb 
and harmonic oscillator systems. However, as shown by Demkov and Ostrosvky 
(Ostrovsky, 2001) in their remarkable work, it is feasible to deduce the 
Madelung rule from an effective one--electron potential of a type similar 
to the one used in the Thomas-Fermi theory of the atom. In some sense, their 
approach exhibits a phenomenological character. However, the result really 
follows from {\it ab initio} calculations in the framework of nonrelativistic 
Quantum Mechanics and, from the mathematical point of view, it corresponds 
to the difficult inverse problem of finding the potential from the spectrum. 
The approach followed in the present work for presenting (not deriving) the 
Madelung rule is entirely different. Indeed, we examine the SO(3) and SO(4) 
content of the rule in order to be prepared to pass to the SO(4,2) and then 
to the SO(4,2)$\otimes$SU(2) format of the periodic table.  

In the central--field approximation, each of the $Z$ electrons of
an atom with atomic number $Z$ is partly characterized by the
quantum numbers $n$, $\ell$, and $m_{\ell}$. The numbers $\ell$
and $m_{\ell}$ are the orbital quantum number and the magnetic
quantum number, respectively. They are connected to the chain of
groups SO(3)$\supset$SO(2): the quantum number $\ell$
characterizes an IRC, of dimension $2\ell+1$, of SO(3) and
$m_{\ell}$ a one--dimensional IRC of SO(2). The principal
quantum number $n$ is such that $n-\ell-1$ is the number of nodes
of the radial wave function associated with the doublet
$(n,\ell)$. In the case of the hydrogen atom or of a
hydrogen--like atom, the number $n$ is connected to the group
SO(4): the quantum number $n$ characterizes an IRC, of dimension
$n^2$, of SO(4). The latter IRC splits into the IRC's
of SO(3) corresponding to $\ell=0,\,1,\,\cdots,\,n-1$ when
SO(4) is restricted to SO(3). A complete characterization of
the dynamical state of each electron is provided by the quartet
$(n,\ell,m_{\ell},m_s)$ or alternatively $(n,\ell,j,m)$. Here, the
spin $s=\frac{1}{2}$ of the electron has been introduced and $m_s$
is the $z$--component of the spin. Furthermore, $j=\frac{1}{2}$
for $\ell=0$ and $j$ can take the values $j=\ell-s$ and $j=\ell+s$
for $\ell\neq0$. The quantum numbers $j$ and $m$ are connected to
the chain of groups SU(2)$\supset$U(1): $j$ characterizes an IRC,
of dimension $2j+1$, of SU(2) and $m$ a one--dimensional IRC of
U(1).

Each doublet $(n,\ell)$ defines an atomic shell. The ground state
of the atom is obtained by distributing the $Z$ electrons of the
atom among the various atomic shells $n\ell$, $n'\ell'$,
$n''\ell''$, $\cdots$ according to (i) an ordering rule and (ii)
the Pauli exclusion principle. A somewhat idealized situation is
provided by the Madelung ordering rule: the energy of the shells
increases with $n+\ell$ and, for a given value of $n+\ell$, with
$n$. This may be depicted by Fig.~1 where the rows are
labelled with $n=1,\,2,\,3,\,\cdots$ and the columns with
$\ell=0,\,1,\,2,\,\cdots$ and where the entry in the $n$--th row
and $\ell$--th column is $[n+\ell,n]$. We thus have the ordering
$[1,1]<[2,2]<[3,2]<[3,3]<[4,3]<[4,4]<[5,3]<[5,4]<[5,5]<[6,4]<[6,5]<[6,6]<\cdots$.
This dictionary order corresponds to the following ordering of the
$n\ell$ shells
$$
1s<2s<2p<3s<3p<4s<3d<4p<5s<4d<5p<6s<\cdots,
$$
which is verified to a good extent by experimental data.

\begin{figure} 
\epsfig{file=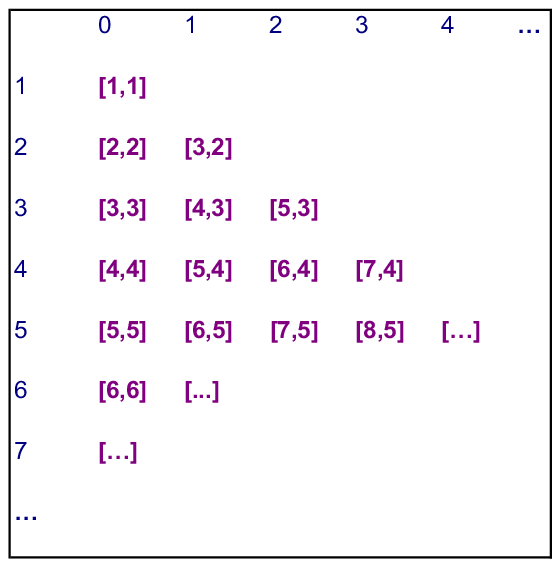}
% \vspace{6pc}
\caption[]{The $[n+\ell,n]$ Madelung array. The lines are labelled
by $n=1,\,2,\,3,\,\cdots$ and the columns by
$\ell=0,\,1,\,2,\,\cdots$. For fixed $n$, the label $\ell$ assumes
the values $\ell=0,\,1,\,\cdots,\,n-1$.}
\label{fig1}
\end{figure}

\begin{figure} 
\epsfig{file=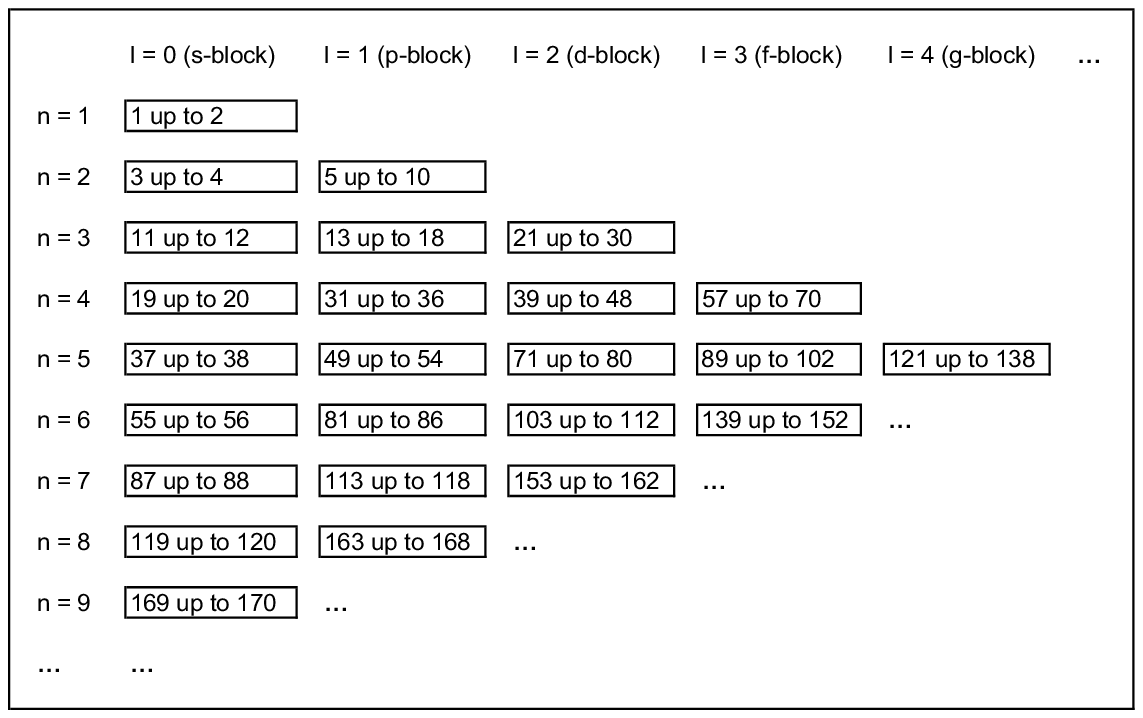}
% \vspace{6pc}
\caption[]{The periodic table deduced from the Madelung array. The
box $[n+\ell,n]$ is filled with $2(2\ell+1)$ elements. The filling
of the various boxes $[n+\ell,n]$ is done according to the
dictionary order implied by Fig.~1.}
\label{fig2}
\end{figure}

From these considerations of an entirely atomic character, we can
construct a periodic table of chemical elements. We start from the
Madelung array of Fig.~1. Here, the significance of the quantum
numbers $n$ and $\ell$ is abandoned. The numbers $n$ and $\ell$
are now simple row and column indexes, respectively. We thus
forget about the significance of the quartet $n$, $\ell$, $j$,
$m$. The various blocks $[n+\ell,n]$ are filled in the dictionary
order, starting from $[1,1]$, with chemical elements of increasing
atomic numbers. More precisely, the block $[n + \ell, n]$ is
filled with $2(2 \ell + 1)$ elements, the atomic numbers of which
increase from left to right. This yields Fig.~2, where each
element is denoted by its atomic number $Z$. For instance, the
block $[1, 1]$ is filled with $2(2 \times 0 + 1)=2$ elements
corresponding to $Z=1$ up to $Z=2$. In a similar way, the blocks
$[2, 2]$ and $[3, 2]$ are filled with $2(2 \times 0 + 1)=2$
elements and $2(2 \times 1 + 1)=6$ elements corresponding to $Z=3$
up to $Z=4$ and to $Z=5$ up to $Z=10$, respectively. It is to be
noted, that the so obtained periodic table {\it a priori} contains
an infinite number of elements: the $n$--th row contains $2n^2$
elements and each column (bounded from top) contains an infinite
number of elements.

\section{The periodic table {\bf\textit{\`a~la}} {SO(4,2)$\otimes$SU(2)}}

We are now in a position to give a group--theoretical articulation
to the periodic table of Fig.~2. For fixed $n$, the $2(2\ell+1)$
elements in the block $[n+\ell, n]$, that we shall refer to an
$\ell$--block, may be labelled in the following way. For $\ell=0$,
the $s$--block in the $n$--th row contains two elements that we
can distinguish by the number $m$ with $m$ ranging from
$-\frac{1}{2}$ to $\frac{1}{2}$ when going from left to right in
the row. For $\ell\neq0$, the $\ell$--block in the $n$--th row can
be divided into two sub-blocks, one corresponding to
$j=\ell-\frac{1}{2}$ (on the left) and the other to
$j=\ell+\frac{1}{2}$ (on the right). Each sub-block contains
$2j+1$ elements, with $2j+1=2\ell$ for $j=\ell- \frac{1}{2}$ and
$2j+1=2(\ell+1)$ for $j=\ell+\frac{1}{2}$, that can be
distinguished by the number $m$ with $m$ ranging from $-j$ to $j$
by step of one unit when going from left to right in the row. In
other words, a chemical element can be located in the table by the
quartet $(n,\ell,j,m)$, where $j=\frac{1}{2}$ for $\ell=0$.

Following Byakov, Kulakov, Rumer and Fet (Konopel'chenko and Rumer, 1979) 
it is perhaps interesting to
use an image with streets, avenues and houses in a city. Let us
call Mendeleev city the city whose west--east streets are
labelled by $n$ and north--south avenues by ($\ell,j,m$). In the
$n$--th street there are $n$ blocks of houses. The $n$ blocks are
labelled by $\ell=0,1,\cdots,n-1$ so that the address of a block
is $(n,\ell)$. Each block contains one sub-block (for $\ell=0$) or
two sub-blocks (for $\ell\neq0$). An address $(n,\ell,j,m)$ can be
given to each house: $n$ indicates the street, $\ell$ the block,
$j$ the sub-block and $m$ the location inside the sub-block. The
organization of the city appears in Fig.~3.

At this stage, it is worthwhile to re-give to the quartet
$(n,\ell,j,m)$ its group--theoretical significance. Then,
Mendeleev city is clearly associated to the IRC
$h\otimes[2]$ of SO(4,2)$\otimes$SU(2) where $[2]$ stands for the
fundamental representation of SU(2). The whole city corresponds
to the IRC
$$
\bigoplus_{n=1}^{\infty}\bigoplus_{\ell=0}^{n-1}
\bigoplus_{j=|\ell-\frac{1}{2}|}^{j=\ell+\frac{1}{2}}(j)=
\left(\bigoplus_{n=1}^{\infty}\bigoplus_{\ell=0}^{n-1}(\ell)\right)\otimes[2]
$$
of SO(4,2)$\otimes$SU(2) in the sense that all the possible
quartets $(n,\ell,j,m)$, or alternatively $(n,\ell,m_{\ell},m_s)$,
can be associated to state vectors spanning this IRC. In the
latter equation, ($\ell$) and ($j$) stand for the IRC's of SO(3)
and SU(2) associated with the quantum numbers $\ell$ and $j$,
respectively.

We can ask the question: How to move in Mendeleev city? Indeed,
there are several bus lines to go from one house to another one?
The SO(3) bus lines, also called 
    SO(3)$\otimes$SU(2) ladder
operators, make it possible to go from one house in a given
$\ell$--block to another house in the same $\ell$--block.
The SO(4) bus lines, also called 
    SO(4)$\otimes$SU(2)
ladder operators, and the SO(2,1) bus lines, also called
SO(2,1) ladder operators, allow to move in a given street 
and in a given avenue, respectively. Finally,
it should be noted that there are taxis, also called
SO(4,2)$\otimes$SU(2) ladder operators, to go from a given house
to an arbitrary house.

Another question concerns the inhabitants, also called chemical
elements, of Mendeleev city. In fact, they are distinguished by a
number $Z$, also called atomic number. The inhabitant living
at the address $(n,\ell,j,m)$ has the number
\begin{eqnarray*}
Z(n\ell jm)&=&\frac{1}{6} (n+\ell)[(n+\ell)^2-1]+\frac{1}{2} (n+\ell+1)^2-\frac{1}{4} [1+(-1)^{n+\ell}]\\
&\times&(n+\ell+1)-4\ell(\ell+1)+\ell+j(2\ell+1)+m-1.
\end{eqnarray*}
Each inhabitant may also have a nickname. All the inhabitants up
to $Z=110$ have a nickname. For example, we have Ds, or
darmstadtium in full, for $Z=110$. Not all the houses in Mendeleev
city are inhabited. The inhabited houses go from $Z=1$ to $Z=116$
(the houses $Z=113$ and $Z=115$ are occupied since the beginning
of 2004). The houses corresponding to $Z\ge117$ are not
presently inhabited. When a house is not inhabited, we also say
that the corresponding element has not been observed yet. The
houses from $Z=111$ to $Z=116$ are inhabited but have not received
a nickname yet. The various inhabitants known at the present time
are indicated on Fig.~4.

It is not forbidden to get married in Mendeleev city. Each
inhabitant may get married with one or several inhabitants
(including one or several clones). For example, we know H$_2$
(including H and its clone), HCl (including H and Cl), and H$_2$O
(including O, H and its clone). However, there is a strict rule in
the city: the assemblages or married inhabitants have to leave the
city. They must live in another city and go to a city sometimes
referred to as a molecular city. Only the clones may stay in
Mendeleev city.

\section{Qualitative aspects of the SO(4,2)$\otimes$SU(2) periodic table}

Going back to Physics and Chemistry, we now describe Mendeleev
city as a periodic table for chemical elements. We have obtained a
table with rows and columns for which the $n$--th row contains
$2n^2$ elements and the $(\ell,j,m)$--th column contains an
infinite number of elements. A given column corresponds to a family 
of chemical analogs, as in the standard periodic table, and a given 
row may contain several periods of the standard periodic table.

The chemical elements in their ground state are considered as
different states of atomic matter: each atom in the table appears
as a particular partner for the (infinite--dimensional) unitary
irreducible representation $h\otimes[2]$ of the group
SO(4,2)$\otimes$SU(2), where SO(4,2) is reminiscent of the
hydrogen atom and SU(2) is introduced for a doubling purpose. In
fact, it is possible to connect two partners of the representation
$h\otimes[2]$ by making use of shift operators of the Lie algebra
of SO(4,2)$\otimes$SU(2). In other words, it is possible to pass
from one atom to another one by means of raising or lowering
operators. The internal dynamics of each element is ignored. In
other words, each neutral atom is assumed to be a noncomposite
physical system. By way of illustration, we give a brief
description of some particular columns and rows of the table.

The alkali--metal atoms are in the first column 
(with $\ell=0$, $j=\frac{1}{2}$, $m=-\frac{1}{2}$, and $n = 1, 2, \cdots$); 
in the atomic shell model, they correspond to an external shell of type $1s$,
$2s$, $3s$, $\cdots$; we note that hydrogen (H) belongs to the
alkali--metal atoms. The second column 
(with $\ell=0$, $j=\frac{1}{2}$, $m= \frac{1}{2}$, and $n = 1, 2, \cdots$) 
concerns the alkaline earth
metals with an external atomic shell of type $1s^2$,
$2s^2$, $3s^2$, $\cdots$; we note that helium (He) belongs to the
alkaline earth metals. The sixth column corresponds to chalcogens 
(with $\ell=1$, $j=\frac{3}{2}$, $m=-\frac{1}{2}$, and $n = 2, 3, \cdots$)
and the seventh column to halogens 
(with $\ell=1$, $j=\frac{3}{2}$, $m= \frac{1}{2}$, and $n = 2, 3, \cdots$);
it is to be observed that hydrogen does not belong to halogens as it
is often the case in usual periodic tables. The eighth column 
(with $\ell=1$, $j=\frac{3}{2}$, $m= \frac{3}{2}$, and $n = 2, 3, \cdots$) 
gives the noble gases with an external atomic shell of type
$2p^6$, $3p^6$, $4p^6$, $\cdots$; helium, with the atomic
configuration $1s^22s^2$, does not belong to the noble gases in
contrast with usual periodic tables.

The $d$--blocks with $n=3$, 4 and 5 yield the three familiar
transition series: 
the iron group goes from Sc(21)
to Zn(30), the palladium group from Y(39) to Cd(48) and the
platinum group from Lu(71) to Hg(80). A fourth transition series
goes from Lr(103) to $Z=112$ (observed but not named yet). In the
shell model, the four transition series correspond to the filling
of the $nd$ shell while the $(n+1)s$ shell is fully occupied, with
$n=3$ (iron group series), $n=4$ (palladium group series), $n=5$
(platinum group series) and $n=6$ (fourth series). The two
familiar inner transition series are the $f$--blocks with $n=4$
and $n=5$:  
the lanthanide series goes from La(57) to
Yb(70) and the actinide series from Ac(89) to No(102). Observe
that lanthanides start with La(57) not Ce(58) and actinides start
with Ac(89) not Th(90). We note that lanthanides and actinides
occupy a natural place in the table and are not reduced to
appendages as it is generally the case in usual periodic tables in
18 columns. A superactinide series is predicted to go from $Z=139$
to $Z=152$ (and not from $Z=122$ to $Z=153$ as predicted by 
G.T. Seaborg). In a shell model approach, the inner transition
series correspond to the filling of the $nf$ shell while the
$(n+2)s$ shell is fully occupied, with $n=4$ (lanthanides), $n=5$
(actinides) and $n=6$ (superactinides). In contrast with Seaborg 
predictions, the table in Fig.~4 shows 
that the elements from $Z=121$ to $Z=138$ form a new period having
no homologue among the known elements.

In Section 5, we have noted that each $\ell$--block with
$\ell\neq0$ gives rise to two sub-blocks. As an example, the
$f$--block for the lanthanides is composed of a sub-block, 
corresponding to $j=\frac{5}{2}$, from La(57) to Sm(62) and
another one, corresponding to $j=\frac{7}{2}$, from Eu(63) to
Yb(70). This division corresponds to the
classification in 
light or ceric  rare earths ($j=\frac{5}{2}$) and 
heavy or yttric rare earths ($j=\frac{7}{2}$). It has received 
justifications both from the experimental (Pascal, 1960) and 
the theoretical (Oudet, 1979) sides.

From a qualitative point of view, the new aspects to come out of this 
theoretical analysis can be summed up as follows: (i) hydrogen and 
helium naturally occur in the first and second columns, respectively; 
(ii) the inner transition series ($d$-block), the transition series 
($f$-block) and the $g$-block occupy a natural place in the table 
(they are not relegated at the periphery of the table); 
(iii) each of the latter blocks (as well as the $p$-block) exhibits 
a division into two sub-blocks that is reminiscent of the relativistic 
splitting 
$(\ell) \to (\ell -–\frac{1}{2}) \oplus (\ell +–\frac{1}{2})$  
which should be significant for heavy elements, cf. the 
distinction between ceric earths and yttric earths (Pascal, 1960); 
(iv) the number of elements afforded by the table is {\it a priori} infinite, 
in view of the infinite--dimensional irreducible representation of SO(4,2) 
on which the table is based (observable elements and/or particles correspond 
to only a few of the allowed quantum mechanical states).

Another radical outcome from this approach concern the possibility of using 
group theory from a quantitative point of view. Chemists are very familiar 
with the use of group--theoretical methods for deriving qualitative results 
(vibration modes, level splitting, selection rules, etc.). We give in the 
following section the main lines of a research programme for quantitatively 
exploiting the potential forces of SO(4,2)$\otimes$SU(2)."

     \begin{figure}
     \epsfig{angle=90,file=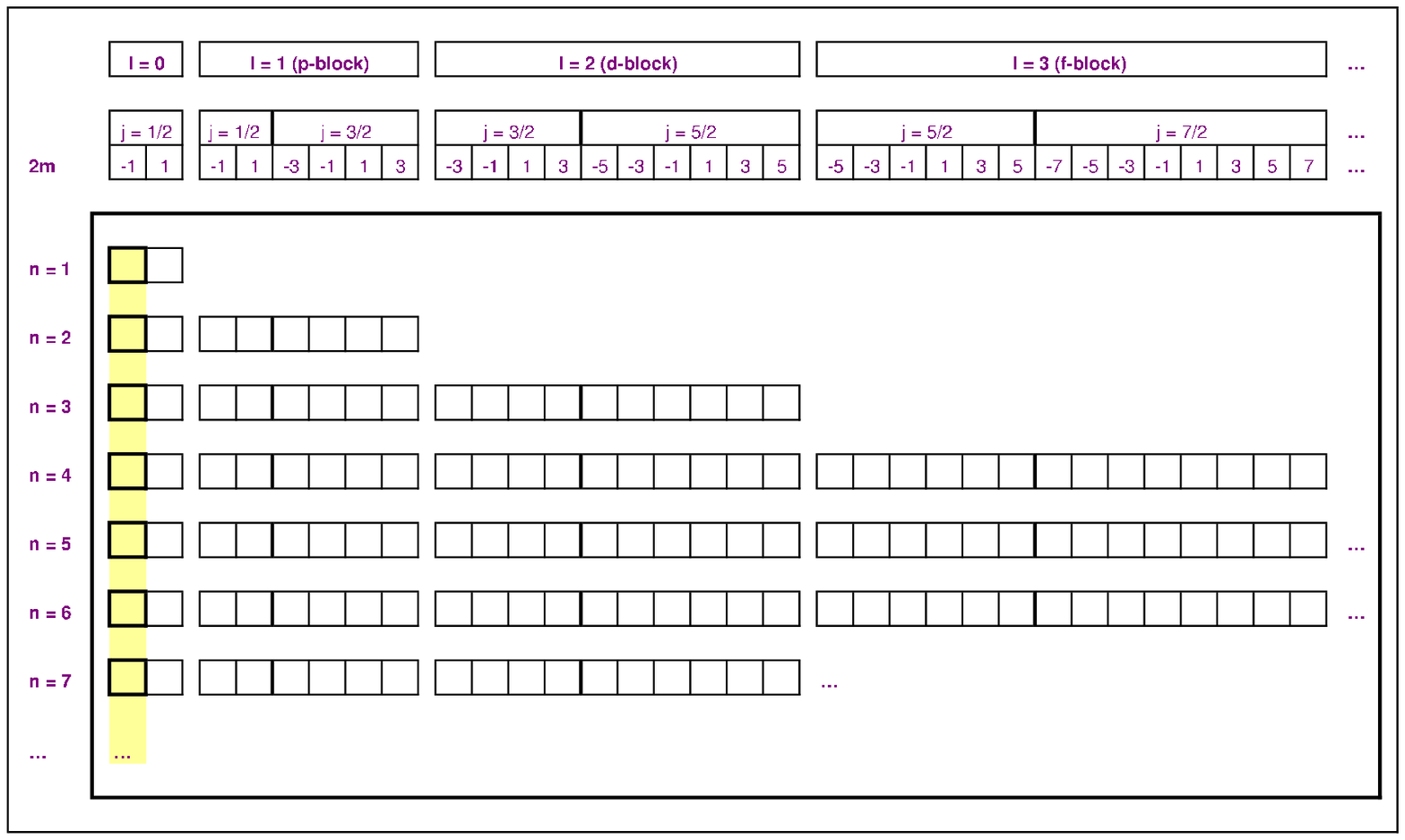}
          %\epsfig{file=KibFig3.eps}
          % \vspace{6pc}
     \caption[]{Mendeleev city. The streets are labelled by
     $n\in\mathbf{N}^*$ and the avenues by $(\ell,j,m)$
     [$\ell=0,1,\cdots,n-1$; $j=\frac{1}{2}$ for $\ell=0$,
     $j=\ell-\frac{1}{2}$ or $j=\ell+\frac{1}{2}$ for $\ell\neq0$;
     $m=-j,-j+1,\cdots,j$].}
     \label{fig3}
     \end{figure}

     \begin{figure}
     \epsfig{angle=90,file=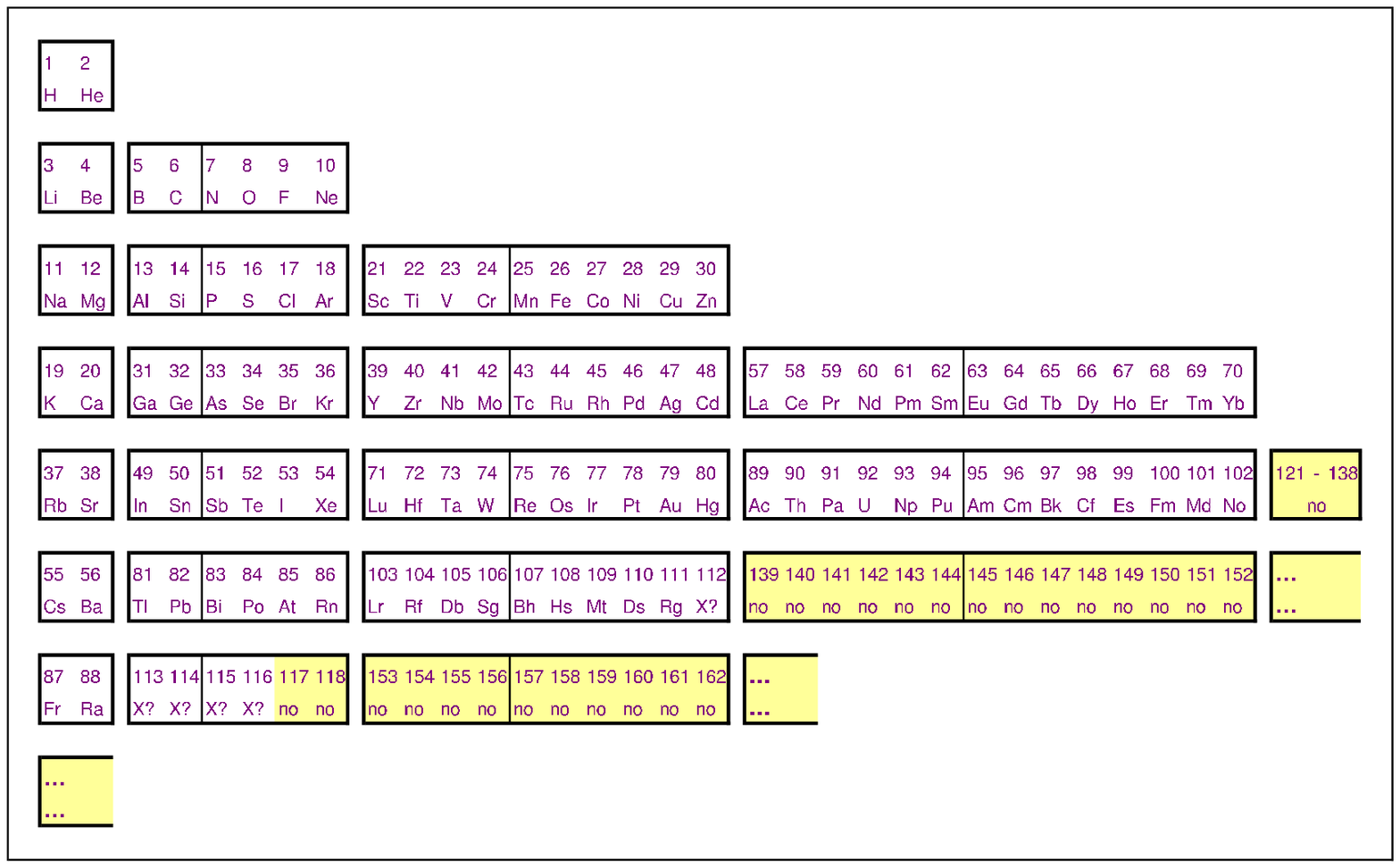}
          %\epsfig{file=KibFig4.eps}
          % \vspace{6pc}
     \caption[]{The inhabitants of Mendeleev city. The houses up to
     number $Z=116$ are inhabited [`X?' means inhabited (or observed)
     but not named, `no' means not inhabited (or not observed)].}
     \label{fig4}
     \end{figure}

\section{Quantitative aspects of the SO(4,2)$\otimes$SU(2) periodic table}

To date, the use of SO(4,2) or SO(4,2)$\otimes$SU(2) in
connection with periodic charts has been limited to qualitative
aspects only, viz., classification of neutral atoms and ions as
well. We would like to give here the main lines of a programme
under development (inherited from nuclear physics and particle
physics) for dealing with quantitative aspects.

The first step concerns the mathematics of the programme. The
direct product group SO(4,2)$\otimes$SU(2) is a Lie group of
order eighteen. Let us first consider the SO(4,2) part which is
a semi-simple Lie group of order $r = 15$ and of rank $\ell=3$. It
has thus fifteen generators involving three Cartan generators
(i.e., generators commuting between themselves). Furthermore, it
has three invariant operators or Casimir operators (i.e.,
independent polynomials, in the enveloping algebra of the Lie
algebra of SO(4,2), that commute with all generators of the
group SO(4,2)). Therefore, we have a set of six ($3+3$)
operators that commute between themselves: the three Cartan
generators and the three Casimir operators. Indeed, this set is
not complete from the mathematical point of view. In other words,
the eigenvalues of the six above-mentioned operators are not
sufficient for labelling the state vectors in the representation
space of SO(4,2). According to a not very well--known result, 
popularized by Racah, we need to find $\frac{1}{2}(r-3\ell)=3$
additional operators in order to complete the set of the six
preceding operators. This yields a complete set of nine ($6+3$)
commuting operators and this solves the state labelling problem
for the group SO(4,2). The consideration of the group SU(2) is
trivial: SU(2) is a semi-simple Lie group of order $r=3$ and of
rank $\ell=1$ so that $\frac{1}{2}(r-3\ell)=0$ in that case. As a
result, we end up with a complete set of eleven ($9+2$) commuting
operators. 

The second step establishes contact with chemical physics. Each of
the eleven operators can be taken to be self-adjoint and thus,
from the quantum--mechanical point of view, can describe an
observable. Indeed, four of the eleven operators, namely, the
three Casimir operators of SO(4,2) and the Casimir operator of
SU(2), serve for labelling the representation $h\otimes[2]$ of
SO(4,2)$\otimes$SU(2) for which the various chemical elements are
partners. The seven remaining operators can thus be used for
describing chemical and physical properties of the elements, as for instance:
ionization energy; oxidation degree; electron affinity;
electronegativity; melting and boiling points; specific heat;
atomic radius; atomic volume; density; magnetic susceptibility;
solubility; etc. In most cases, this can be done by expressing a
chemical observable associated with a given property 
(for which we have few experimental data) in terms of
the seven operators which serve as an integrity basis for the
various observables. Each observable can be developed as a linear
combination of operators constructed from the integrity basis.
This is reminiscent of group--theoretical techniques used in
nuclear and atomic spectroscopy (cf. the Interacting Boson Model)
or in hadronic spectroscopy (cf. the Gell-Mann/Okubo mass formulas
for baryons and mesons).

The last step is to proceed with a diagonalization process and then
to fit the various linear combinations to experimental data. This
can be achieved through fitting procedures concerning either a
period of elements, taken along a same line of the periodic table, 
or a family of elements, taken along a same column of the periodic
table. For each property this will lead to a formula or
phenomenological law that can be used in turn for making
predictions concerning the chemical elements for which no data are
available. In addition, it is hoped that this will shed light on
regularities and well--known as well as recently discovered patterns 
of the periodic table, such as unexpected patterns connecting elements 
via a knight's move in the table (Laing, 2004; Rayner--Canham, 2004)."

\section{Closing remarks}

Group--theoretical methods based on symmetry considerations have been 
continuously developed during the 20th century in order to classify the 
constituents of matter and to understand their interactions. The 
SO(4,2)$\otimes$SU(2) periodic table presented in this article was set up along 
lines similar to the ones used for classifying fundamental particles via flavor 
groups. The group SO(4,2)$\otimes$SU(2) is a flavor group in the sense that 
each chemical element appears to be a particular state (or flavor) of a single 
element.

We close this paper with two remarks. Possible extensions of the
work presented in Sections 5--7 concern isotopes 
and molecules. The consideration of isotopes 
needs the introduction of the number of nucleons in the atomic
nucleus. With such an introduction we have to consider other
dimensions for Mendeleev city: the city is no longer restricted
to spread in Flatland. Group--theoretical analyses of periodic
systems of molecules can be achieved by considering direct
products involving several copies of SO(4,2)$\otimes$SU(2).
Several works have been already devoted to this subject 
(Kibler, 2006).

\section*{Acknowledgements}

Thanks are due to the Referee for pertinent criticism and useful suggestions.


\begin{thebibliography}{99}

\bibitem{Barut}
Barut, A.O.
    {Group Structure of the Periodic System}.
    In B.G. Wybourne, editor, 
{\em The Structure of Matter}. University of Canterbury 
Publications, Christchurch, New Zealand, pp. 126-136, 1972.

\bibitem{Barut-Kleinert}
Barut, A.O. and H. Kleinert.
    {Transition Form Factors in the H Atom}.
    {\em Phys. Rev.}, 160:1149--1151, 1967.

\bibitem{Kibler}
Kibler, M.R.
    {On the Use of the Group SO(4,2) in Atomic and Molecular Physics}.
    {\em Mol. Phys.}, 102:1221--1230, 2004.

\bibitem{Kibler}
Kibler, M.R.
    {A group--Theoretical Approach to the Periodic Table: Old and New
Developments}.
    In D.H. Rouvray and R.B. King, editors, 
{\em The Mathematics of the Periodic Table}. Nova Science, NY, pp. 237-263, 2006.

\bibitem{Kibler-N}
Kibler, M. and T.~N\'egadi.
    {Connection between the Hydrogen Atom and the Harmonic Oscillator}.
    {\em Phys. Rev. A}, 
29:2891--2894, 1984.

\bibitem{Konopel'chenko-Rumer}
Konopel'chenko, V.G. and Yu.B. Rumer.
    {Atoms and Hadrons (Classification Problems)}.
    {\em Sov. Phys. Usp.}, 
22:837--840, 1979.

\bibitem{Laing}
Laing, M. 
Patterns in the Periodic Table -- Old and New. 
In D.H. Rouvray and R.B. King, editors, 
{\em The Periodic Table: Into the 21st Century}. 
Research Studies Press, Baldock, U.K., pp. 123-140, 2004.

\bibitem{Lowdin}
L\"owdin, P.-O. 
Some Comments on the Periodic System of the Elements. 
{\em Int. J. Quantum Chem., Quantum Chem. Symp.}, 3:331-334, 1969.

\bibitem{Malkin-Man'ko}
Malkin, I.A. and V.I. Man'ko.
    {Symmetry of the Hydrogen Atom}.
    {\em Sov. Phys. JETP Lett.}, 2:146--148, 1965.

\bibitem{Ostrovsky}
Ostrovsky, V.N. 
What and How Physics Contributes to Understanding the 
Periodic Law. {\em Found. Chem.}, 3:145-182, 2001.

\bibitem{Oudet}
Oudet, X. 
Valency, Ionicity and Electronic Configuration in Rare Earths. 
{\em J. Physique (Paris)}, 40(C5):395-397, 1979.

\bibitem{Pascal}
Pascal, P. 
{\em Nouveau trait\'e de chimie min\'erale}. Masson, Paris, 1960. 

\bibitem{Rayner-Canham}
Rayner-Canham, G.W. 
The Richness of Periodic Patterns. 
In D.H. Rouvray and R.B. King, editors, 
{\em The Periodic Table: Into the 21st Century}. 
Research Studies Press, Baldock, U.K., pp. 161-186, 2004.

\bibitem{Rouvray-King}
Rouvray, D.H. and R.B. King.
    {\em The Periodic Table: Into the 21st Century}.
    Research Studies Press, Baldock, U.K., 2004.

\bibitem{Rumer-Fet}
Rumer, Yu.B. and A.I. Fet.
    {The Group Spin(4) and the Mendeleev System}.
    {\em Teoret. Mat. Fiz.}, 9:203--210, 1971.

\bibitem{Scerri2}
Scerri, E. 
Prediction of the Nature of Hafnium from Chemistry, Bohr's Theory 
and Quantum Theory. {\em Ann. Sc.}, 51:137-150, 1994.

\bibitem{Scerri3}
Scerri, E. 
Commentary on Allen \& Knight's Response to the L\"owdin Challenge. 
{\em Found. Chem.}, 8:285-293, 2006.

\bibitem{Scerri}
Scerri, E.
    {\em The Periodic Table: Its Story and Its Significance}.
    Oxford University Press, New York, NY, 2007.

\bibitem{Spronsen}
van Spronsen, J.W.
    {\em The Periodic System of Chemical Elements: A History of the First
Hundred Years}.
    Elsevier, Amsterdam, 1969.

\end{thebibliography}
\end{document}